\documentclass[a4paper,8pt]{article} 
\usepackage{m-floating}
\usepackage{graphicx}
\usepackage{texab}
\usepackage{amsmath,amssymb}
\usepackage{amsthm}
\usepackage{bm}
\usepackage{natbib}
\usepackage{algorithm}
\usepackage{algorithmic}
\usepackage{indentfirst}

\begin{document}

\bibliographystyle{apalike}

\title{GLMMLasso: An Algorithm for High-Dimensional Generalized Linear
  Mixed Models Using $\ell_1$-Penalization}
\author{J\"urg Schelldorfer$^1$,  Lukas Meier$^2$ and Peter B\"uhlmann$^2$  \thanks{juerg.schelldorfer@axa-winterthur.ch, meier@stat.math.ethz.ch, buhlman@stat.math.ethz.ch} \vspace{0.5cm} \\ $^1$AXA Winterthur\\ 
$^2$Seminar f\"ur Statistik, ETH Z\"urich }


\maketitle

\begin{abstract}
  We propose an $\ell_1$-penalized algorithm for fitting high-dimensional
  generalized linear mixed models. Generalized linear mixed models (GLMMs)
  can be viewed as an extension of generalized linear models for clustered
  observations. Our Lasso-type approach for GLMMs should be mainly used as
  variable screening method to reduce the number of variables below the
  sample size. We then suggest a refitting by maximum likelihood based on
  the selected variables only. This is an effective correction to overcome
  problems stemming from the variable screening procedure which are more
  severe with GLMMs than for generalized linear models. We illustrate the
  performance of our algorithm on simulated as well as on real data
  examples. Supplemental materials are available online and the algorithm
  is implemented in the \texttt{R} package \texttt{glmmixedlasso}.
\end{abstract}

{\bf Key Words:} {coordinate gradient descent; Laplace approximation;
  random-effects model;\\
 variable selection.}\\

\section{Introduction}
\indent In recent years, high-dimensional linear regression models have
been extensively studied. The most popular method to achieve sparse
estimates is the Lasso \citep{Tibs96}, which uses an $\ell_1$-penalty. The
Lasso is not only attractive in terms of its statistical properties but
also due to its fast computation solving a convex optimization problem.
However, relatively few articles examine high-dimensional regression
problems involving a non-convex loss function, i.e.\ \cite{Khalili07} and
\cite{Stad10} for Gaussian mixture
models, \cite{Pan07} and \cite{Witten10} for clustering and \cite{Witten11} for linear discriminant analysis.\\
\indent Generalized linear mixed models
\citep{McCPN89,Breslow93,McCCS01,MolGV05} are an extension of generalized
linear models by adding random effects to the linear predictor in order to
accommodate for clustered or overdispersed data.  These models have received
much attention in many applications such as biology, ecology, medicine,
pharmaceutical science and econometrics. Available software packages
(\texttt{lme4} in \texttt{R}, NLMIXED in SAS, among
others) allow to fit a wide range of generalized linear mixed models.\\
\indent In this paper we develop a method for high-dimensional generalized
linear mixed models. It is based on a Lasso-type regularization with a
cyclic coordinate descent optimization. Due to shrinkage introduced by
$\ell_1$-penalization, our approach performs in a first step variable
screening, thereby selecting a set of candidate active variables. In other
words, the proposed method primarily aims at reducing the dimensionality of
the high-dimensional GLMM. In a second step, we perform refitting by
maximum likelihood estimation to get accurate parameter estimates. The idea
of such a two-stage approach has been used in linear models \citep{Efro04}
and it is related to the adaptive Lasso \citep{Zou06} and the thresholded
Lasso \citep{Zhou10,GeerBuhlZhou10}. In fact, a two-stage approach is much
more important than for linear models since shrinkage in GLMMs can have a
severe effect on the estimation of
variance components, see Sections \ref{Sec4} and \ref{Sec5}.\\
\indent To the best of our knowledge, there does not exist any literature
devoted to truly high-dimensional generalized linear mixed models. Some
papers focus on penalized variable selection procedures in generalized
mixed models with low-dimensional data: we refer to \cite{Yang07},
\cite{Ibrahim10}, \cite{Ni10}. \cite{Groll11} have independently studied
the same statistical problem and have also used a Lasso-type approach but
with a focus on rather low-dimensional problems. Few papers focus on
variable selection in generalized additive mixed models, for example
\cite{Xue10} and \cite{Lai12}. \cite{Schell11} present statistical theory
and an algorithm for high-dimensional Gaussian linear mixed models, where
computation is much easier than in the generalized case.

The main contribution of the present paper is the construction and
implementation of an efficient algorithm for $\ell_1$-penalization in truly
high-dimensional generalized linear mixed models, called the GLMMLasso. We
use the Laplace approximation \citep{Bates09b} and combine it with
efficient coordinate gradient descent methods \citep{Tseng09}. Our
algorithm is feasible for problems where the number of variables is in the
thousands and taking advantage of sparsity with respect to dimensionality
(i.e.\ only
few active variables) is exploited by an active set strategy.\\
\indent The rest of the article is organised as follows. In Section
\ref{Sec2}, we review the generalized linear mixed model and introduce the
GLMMLasso estimator. In Section \ref{Sec3}, we describe the details of the
computational algorithm before advocating the two-stage GLMMLasso
estimators in Section \ref{Sec4}. In Section \ref{Sec5} and \ref{Sec6} we
consider the performance of our methods on simulated and real data
sets. The article concludes with a discussion in Section
\ref{Sec7}. Supplemental materials including additional simulation examples
are available online. 

\section{Generalized linear mixed models and $\ell_1$-penalized
  estimation}\label{Sec2} 
\indent In this section, we first look at the classical GLMM setting where
the number of observations is larger than the number of covariates,
i.e.\ $p<n$. We closely follow \cite{Bates09c}. Secondly, we consider the
high-dimensional framework, i.e.\ $n \ll p$, and present the
$\ell_1$-penalized maximum likelihood estimator.

\subsection{Model formulation}
Suppose that the observations are not independent but grouped instead. Let
$r=1,\ldots,N$ be the grouping index and $j=1,\ldots,n_r$ the $j$th outcome
within group $r$. Denote by $n$ the total number of observations,
i.e.\ $n=\sum_{r=1}^N n_r$. Let $\bm{X}$ be the $n \times p$ fixed-effects
design matrix, $\bm{Z}$ the $n \times q$ random-effects design matrix,
$\bm{\mathcal{Y}}$ the $n$-dimensional random response vector and
$\bm{\mathcal{B}}$ be the $q$-dimensional vector of random effects. We
observe $\bm{y}$ of $\bm{\mathcal{Y}}$ whereas $\bm{\mathcal{B}}$ is
unobserved. The generalized linear mixed model is specified by the
unconditional distribution of $\bm{\mathcal{B}}$ and the conditional
distribution of $\bm{\mathcal{Y}}|\bm{\mathcal{B}}=\bm{b}$:
\begin{itemize}
\item [i)]$\mathcal{Y}_i|\bm{\mathcal{B}}=\bm{b}$ are independent for
  $i=1,\ldots,n$.
\item [ii)] The distribution of $\mathcal{Y}_i|\bm{\mathcal{B}}=\bm{b}$
  belongs to the exponential family with
  density $$\exp \Big\{\phi^{-1}\Big(y_{i}\xi_i - b(\xi_i)\Big) +
  c(y_i,\phi)\Big\}, $$ where $b(.)$ and $c(.,.)$ are known
  functions. $\phi$ is the dispersion parameter (known or unknown) and $\xi_i$
  is associated with the conditional mean
  $\mu_i:=E[\mathcal{Y}_i|\bm{\mathcal{B}}=\bm{b}]$, i.e.\ $\xi_i=\xi_i(\mu_i)$.
\item [iii)] The conditional mean vector $\bm{\mu}$ depends on $\bm{b}$
  through the known link function $g$ and the linear predictor
  $\bm\eta=\bm{X\beta} + \bm{Zb}$, with $\bm\eta=\bm{g}(\bm\mu)$
  componentwise. Here, $\bm\beta$ is the unknown $p$-dimensional parameter
  vector, called fixed effects, and $\bm{b}$ the unknown $q$-dimensional
  vector of random effects.
\item [iv)] $\bm{\mathcal{B}} \sim \mathcal{N}_q(\bm{0},\bm\Sigma_{\bm{\theta}})$
  where the covariance matrix $\bm\Sigma_{\bm{\theta}}$ is parameterized by
  the unknown parameter vector $\bm{\theta} \in \mathbb{R}^d$. We assume that
  $\bm\Sigma_{\bm\theta}$ is positive semidefinite, i.e.\ $\bm\Sigma_{\bm\theta}
  \ge 0$. The dimensionality $d$ is typically small, say $d \le 10$.
\end{itemize}
By using $\bm{\mathcal{B}}$ and $\bm\Sigma_{\bm\theta}$ in the definition
above, we have already defined the random-effects structure of the GLMM. To
be more precise, we have specified which variables have an additional
random effect and how the structure of $\bm\Sigma_{\bm\theta}$ looks like
(e.g.\ multiple of the identity or diagonal). A discussion of how to find
these structures is beyond
the scope of this paper.\\
\indent Let us write $\bm\Sigma_{\bm\theta}$ in terms of its Cholesky
decomposition
$\bm\Sigma_{\bm\theta}=\bm\Lambda_{\bm\theta}\bm\Lambda_{\bm\theta}^T$ and
introduce the (unobserved) random variable $\bm{\mathcal{U}}$ defined by
$\bm{\mathcal{B}}:=\bm\Lambda_{\bm\theta}\bm{\mathcal{U}}$ where
$\bm{\mathcal{U}} \sim \mathcal{N}_q(\bm{0},\bm{1}_q)$. Then the linear
predictor $\bm{\eta}$ can be written as $\bm\eta=\bm{X}\bm\beta +
\bm{Z}\bm\Lambda_{\bm\theta}\bm{u}$. We estimate the parameters
$\bm{\beta}$, $\bm{\theta}$ and $\phi$ (if unknown) by the maximum
likelihood method and predict the random effects $\bm{u}$.

\subsection{Likelihood function}
Employing the notation $\xi_i(\mu_i)=\xi_i(\bm\beta,\bm\theta)$,
the likelihood function of a GLMM is given by the following expression:
\small
\begin{align} \label{llgeneral}
  L(\bm\beta,\bm\theta,\phi)&=\int_{\mathbb{R}^q} \prod_{i=1}^n \Bigg[\exp
  \Bigg\{\phi^{-1}\Big(y_{i}\xi_i(\bm\beta,\bm\theta) -
  b(\xi_i(\bm\beta,\bm\theta))\Big) + c(y_i,\phi) \Bigg\}
  \Bigg]\frac{1}{(2\pi)^{q/2}} \exp \Big\{-\frac{1}{2}\|\bm{u}\|_2^2
  \Big\} d\bm{u} \nonumber \\
  &= \frac{1}{(2\pi)^{q/2}} \int_{\mathbb{R}^q} \exp \Bigg\{\sum_{i=1}^n
  \Big(\frac{y_{i}\xi_i(\bm\beta,\bm\theta) -
    b(\xi_i(\bm\beta,\bm\theta))}{\phi} + c(y_i,\phi)\Big)
  -\frac{1}{2}\|\bm{u}\|_2^2 \Bigg\} d\bm{u}.
\end{align}
\normalsize In general, the integral (\ref{llgeneral}) can not be worked
out analytically and numerical approximations are required, see
\cite{Skrondal04}, \cite{MolGV05} and \cite{Jiang07}.

\subsection{The GLMMLasso estimator}
We now turn to the high-dimensional setting where the number of
fixed-effect variables $p$ is much larger than the number of observations
$n$, i.e.\ we study the so-called $n \ll p$ setup.\\
\indent Let us assume that the true underlying fixed-effects vector
$\bm{\beta}_0$ is sparse in the sense that many coefficients of
$\bm{\beta}_0$ are zero. To enforce sparsity of our estimator, we advocate
a Lasso-type approach. This means that we add an $\ell_1$-penalty for the
fixed-effects vector $\bm\beta$ to the likelihood function. Thus, we are
going to consider the following objective function:
\begin{equation} \label{Qfunction}
Q_{\lambda}(\bm\beta,\bm\theta,\phi)= -2\log L(\bm\beta,\bm\theta,\phi) +
\lambda\|\bm{\beta}\|_1, 
\end{equation}
where $\lambda \ge 0$ is a regularization parameter. Appropriate choices
for
$\lambda$ are discussed in Section \ref{Sec4}.\\
\indent We aim at estimating the fixed-effect parameter $\bm\beta$, the
covariance parameter $\bm\theta$, and if unknown the dispersion parameter
$\phi$, by
\begin{equation} \label{obfct1}
(\hat{\bm\beta},\hat{\bm{\theta}},\hat{\phi}) := \argmin_{\bm\beta,\bm\theta,\phi} Q_{\lambda}(\bm\beta,\bm\theta,\phi).
\end{equation}
We call (\ref{obfct1}) the GLMMLasso estimator. Since the likelihood
function (\ref{llgeneral}) comprises analytically intractable integrals
(except for the Gaussian case), some approximations have to be used. We are
going to illustrate the algorithm using the Laplace approximation. For
GLMMs, it is accurate with low computational burden, as advocated by
\cite{Bates09b}. A thorough discussion of the accuracy and limitations of
the Laplace approximation can be found in \cite{Joe08}. Generally, the
Laplace approximation is used to calculate integrals of the form
\begin{equation}
I = \int_{\mathbb{R}^q} e^{-S(\bm{u})}d\bm{u},
\end{equation}
where $S(\bm{u})$ is a known function of a
$q$-dimensional variable $\bm{u}$. Let
\begin{equation} \label{mode}
\tilde{\bm{u}}=\argmax_{\bm{u}}-S(\bm{u})
\end{equation}
(i.e.\ $S'(\tilde{\bm{u}})=0$), then the Laplace approximation of $I$ is given by
\begin{equation}\label{laplaceapprox}
I \approx I^{LA} = (2 \pi)^{q/2} |S''(\tilde{\bm{u}})|^{-1/2}
e^{-S(\tilde{\bm{u}})}.
\end{equation}
The mode $\bm{\tilde{u}}$ in (\ref{mode}) is calculated by the penalized
iterative least squares (PIRLS) algorithm. It is presented in
\cite{Bates09b} and described in the supplemental materials. The PIRLS
algorithm is related to the iterative reweighted least squares (IRLS)
algorithm for obtaining the maximum
likelihood estimator in generalized linear models.\\
\indent It should be noted that $\bm{\tilde{u}}$ depends on $\bm{\beta}$,
$\bm{\theta}$ and $\phi$. From (\ref{llgeneral}) and (\ref{laplaceapprox})
we deduce that the Laplace approximation of the objective function
$Q_{\lambda}(.)$ in (\ref{Qfunction}) is
\begin{align} \label{laplaceobfct}
Q_{\lambda}^{LA}(\bm\beta,\bm\theta,\phi) =& -2 \sum_{i=1}^n \Bigg\{ \frac{y_{i}\xi_i(\bm\beta,\bm\theta) - b(\xi_i(\bm\beta,\bm\theta))
}{\phi} + c(y_i,\phi)
\Bigg\} + \log|(\bm{Z}\bm\Lambda_{\bm\theta})^T\bm{W}_{\bm\beta,\bm\theta,\phi}
(\bm{Z}\bm\Lambda_{\bm\theta}) + \bm{1}_q| \\ \nonumber
& + \|\tilde{\bm{u}}(\bm\beta,\bm\theta,\phi)\|_2^2  +
\lambda \|\bm\beta\|_1, 
\end{align}
where $\bm{W}_{\bm\beta,\bm\theta,\phi}=\diag^{-1}\Big(\phi v(\mu_i(\bm\beta,\bm\theta))g'(\mu_i(\bm\beta,\bm\theta))^2\Big)_{i=1}^n$ and
$v(.)$ is the known conditional variance function \citep{McCPN89}. The estimator
(\ref{obfct1}) is then approximated by
\begin{equation} \label{laobfct}
(\hat{\bm\beta}^{LA},\hat{\bm{\theta}}^{LA},\hat{\phi}^{LA}) := \argmin_{\bm\beta,\bm\theta,\phi}
Q^{LA}_{\lambda}(\bm\beta,\bm\theta,\phi).
\end{equation}
We call (\ref{laobfct}) the GLMMLasso$^{LA}$ estimator. It is the approximation
(\ref{laobfct}) to the objective function (\ref{obfct1}) that is optimized
to obtain the parameter estimates. Moreover, we would like to emphasize
that (\ref{laobfct}) is a non-convex function with respect to
$(\bm\beta,\bm\theta,\phi)$ consisting of a non-convex loss function and a convex penalty.

\section{Computational algorithm} \label{Sec3}
In this section, we present the computational algorithm to obtain
the GLMMLasso$^{LA}$ estimator (\ref{laobfct}). The algorithm is based on ideas from \cite{Tseng09} of the (block)
coordinate gradient descent (CGD) method. The notion of the CGD
algorithm is that we cycle through components of the full parameter vector
$\bm\psi:=(\bm\beta,\bm\theta,\phi) \in \mathbb{R}^{p+d+1}$ and
minimize the objective function $Q^{LA}_{\lambda}(.)$ only with respect to one
parameter while keeping the other parameters fixed. In doing so we calculate a
quadratic approximation and perform an indirect line search to ensure that
the objective function decreases. (Block) CGD algorithms are used in
\cite{Meie08}, \cite{Wu08}, \cite{Fried08} and \cite{Breheny11} and are now
extremely popular
in high-dimensional penalized regression problems.\\
\indent We first give an overview of the algorithm which solves
minimization problem (\ref{laobfct}) exactly 
before considering an approximate algorithm which finds a solution close to
the exact minimizer of (\ref{laobfct}). Finally, we present some details of
the algorithm.

\subsection{The exact GLMMLasso algorithm}
We describe here an exact algorithm, called exact GLMMLasso (we notationally omit
the involved Laplace approximation), for the Laplace
approximated objective function in (\ref{laobfct}). Let us write
(\ref{laplaceobfct}) with a different notation to ease the
presentation. For $\bm\psi=(\bm\beta,\bm\theta,\phi) \in
\mathbb{R}^{p+d+1}$, define the function 
\begin{equation*}
f(\bm{\psi}) := -2 \sum_{i=1}^n \Bigg\{ \frac{y_{i}\xi_i(\bm\beta,\bm\theta) - b(\xi_i(\bm\beta,\bm\theta))
}{\phi} + c(y_i,\phi)
\Bigg\} + \log|(\bm{Z}\bm\Lambda_{\bm\theta})^T\bm{W}_{\bm\psi}
(\bm{Z}\bm\Lambda_{\bm\theta}) + \bm{1}_q| + \|\tilde{\bm{u}}(\bm\psi)\|_2^2.
\end{equation*}
Now (\ref{laobfct}) can be written as $\hat{\bm{\psi}}_{\lambda}^{LA} =
\argmin_{\bm{\psi}} Q_{\lambda}^{LA}(\bm{\psi}) := f(\bm{\psi}) +
\lambda \|\bm\beta \|_1$. Let $\bm{e}_j$ be the $j$th unit vector and
denote by $(s)$ the $s$th iteration step. Moreover, we let
\begin{equation*} \nonumber
\bm\beta^{(s)}:= (\beta_1^{(s)},\ldots,\beta_p^{(s)})^T , \quad
\bm\theta^{(s)}:=(\theta_1^{(s)},\ldots,\theta_d^{(s)})^T, \quad \phi^{(s)}
\end{equation*}
be the estimates of $\bm\beta$, $\bm\theta$ and $\phi$ in the $s$th
iteration. Using the notation
\begin{align} \nonumber
\bm\beta^{(s,s-1,\beta_k)}&:=
\Big(\beta_1^{(s)},\ldots,\beta_{k-1}^{(s)},\beta_k,\beta_{k+1}^{(s-1)},\ldots,\beta_{p}^{(s-1)} 
\Big)^T,\\  \nonumber
\bm\theta^{(s,s-1,\theta_l)} &:=
\Big(\theta_1^{(s)},\ldots,\theta_{l-1}^{(s)},\theta_{l},\theta_{l+1}^{(s-1)},\ldots,\theta_d^{(s-1)}\Big)^T,\\  \nonumber
\bm\beta^{(s,s-1;k)}&:=
\Big(\beta_1^{(s)},\ldots,\beta_{k-1}^{(s)},\beta_k^{(s-1)},\beta_{k+1}^{(s-1)},\ldots,\beta_{p}^{(s-1)}
\Big)^T,
\end{align}
the exact GLMMLasso algorithm is summarized in Algorithm \ref{alg1}.

\begin{algorithm} [!h]
\caption{\textit{Exact GLMMLasso algorithm}}
\label{alg1}
\footnotesize
\begin{enumerate}
\item [(0)] Choose a starting value
  $\bm\psi^{(0)}=(\bm\beta^{(0)},\bm\theta^{(0)},\phi^{(0)})$. 
\end{enumerate}
\noindent \textbf{Repeat} for $s=1,2,\ldots$
\begin{enumerate}
\item [(1)] \emph{(fixed-effect parameter optimization)}\\ For $k=1,\ldots,p$
\begin{enumerate}
\item [a)] \emph{(Laplace approximation)}\\
Calculate the Laplace approximation
  $$Q_{\lambda}^{LA}\Big(\bm\beta^{(s,s-1;k)},\bm\theta^{(s-1)},\phi^{(s-1)}\Big).$$
\item [b)] \emph{(Quadratic approximation and inexact line search)}
\begin{itemize}
\item [i)] Approximate the second derivative
 $$\frac{\partial^2}{\partial
    \beta_k^2}
    f\Big(\bm\beta^{(s,s-1,\beta_k)},\bm\theta^{(s-1)},\phi^{(s-1)}\Big)\Big|_{\beta_k=\beta_k^{(s-1)}}$$
    by 
  $h_k^{(s)} > 0$ as described in the subsection below.
\item [ii)]  Calculate the descent direction $d_{k}^{(s)} \in \mathbb{R}$
\begin{align*}
d_k^{(s)} :=& \argmin_d  \Big\{
f\Big(\bm\beta^{(s,s-1;k)},\bm\theta^{(s-1)},\phi^{(s-1)}\Big) + 
\frac{\partial}{\partial \beta_k}
f\Big(\bm\beta^{(s,s-1,\beta_k)},\bm\theta^{(s-1)},\phi^{(s-1)}\Big)\Big|_{\beta_k=\beta_k^{(s-1)}}d\\ 
& + \frac{1}{2} d^2 h_k^{(s)} +
  \lambda \|\bm{\beta}^{(s,s-1;k)} + d \bm{e}_{k}\|_1 \Big\}.
\end{align*}
\item [iii)] Choose a step size $\alpha_k^{(s)} >0$  and set
  $\bm{\beta}^{(s,s-1;k+1)} = \bm{\beta}^{(s,s-1;k)} + \alpha^{(s)}_k d_k^{(s)}
  \bm{e}_k$ such
  that $$Q_{\lambda}^{LA}\Big(\bm\beta^{(s,s-1;k+1)},\bm\theta^{(s-1)},\phi^{(s-1)}\Big)
  \le Q_{\lambda}^{LA}\Big(\bm\beta^{(s,s-1;k)},\bm\theta^{(s-1)},\phi^{(s-1)}\Big).$$
\end{itemize}
\end{enumerate}
\item [(2)] \emph{(Covariance parameter optimization)}\\ For $l=1,\ldots,d$
\begin{equation*}
\theta_{l}^{(s)} = \argmin_{\theta_l}
Q^{LA}_{\lambda}\Big(\bm\beta^{(s)},\bm\theta^{(s,s-1;\theta_l)},\phi^{(s-1)}\Big).
\end{equation*}
\item [(3)] \emph{(Dispersion parameter optimization)}
\begin{equation*}
\phi^{(s)} = \argmin_{\phi}
Q^{LA}_{\lambda}\Big(\bm\beta^{(s)},\bm\theta^{(s)},\phi \Big).
\end{equation*}
\end{enumerate}
\noindent \textbf{until} convergence.
\end{algorithm}
\normalsize
Particularly in the high-dimensional setting, the
calculation of the quadratic approximation requires a large amount of
computing time. Therefore it is interesting to examine a much faster approximate algorithm.

\subsection{The (approximate) GLMMLasso algorithm}
In the exact Algorithm \ref{alg1} above, we consider in step (1) b)
the mode $\bm{\tilde{u}}$ as a function of the parameters, i.e.\
$\tilde{\bm{u}}=\tilde{\bm{u}}(\bm\beta,\bm\theta,\phi)$. However, the
calculation of the derivatives of $f(.)$ with respect to $\beta_k$ is
computationally intensive. This becomes a major issue in the
high-dimensional setting where a substantial amount of computing time is
allocated to this particular part of the algorithm. In addition, the exact
GLMMLasso algorithm 
requires a large number of outer iterations $s$. To attenuate these
difficulties, we propose a slightly modified version of Algorithm
\ref{alg1}. We suggest performing the quadratic approximation and the
inexact line search while considering $\bm{\tilde{u}}$ as
fixed and not depending on $\beta_k$. Denoting by $f(.|\tilde{\bm{u}})$ the
function $f(.)$ for which $\tilde{\bm{u}}$ is considered as fixed, the
(approximate) GLMMLasso algorithm is given in Algorithm \ref{alg2}:

\begin{algorithm} [!h]
\caption{\textit{(Approximate) GLMMLasso algorithm}}
\label{alg2}
\footnotesize
Denote by
$\tilde{\bm{u}}=\tilde{\bm{u}}\big(\bm\beta^{(s,s-1;k)},\bm\theta^{(s-1)},\phi^{(s-1)}
\big) $. Replace in Algorithm \ref{alg1} i) - iii) by
\begin{itemize}
\item [i')] Approximate the second derivative
 $$\frac{\partial^2}{\partial
    \beta_k^2}
    f\Big(\bm\beta^{(s,s-1,\beta_k)},\bm\theta^{(s-1)},\phi^{(s-1)}\Big|\tilde{\bm{u}}\Big)\Big|_{\beta_k=\beta_k^{(s-1)}}$$ 
    by 
  $h_k^{(s)} > 0$ as described in the subsection below.
\item [ii')] Calculate the descent direction $d_{k}^{(s)} \in \mathbb{R}$
\begin{align*}
d_k^{(s)} :=& \argmin_d  \Big\{ f\Big(\bm\beta^{(s,s-1;k)},\bm\theta^{(s-1)},\phi^{(s-1)}\Big|\tilde{\bm{u}}\Big) +
\frac{\partial}{\partial \beta_k}
f\Big(\bm\beta^{(s,s-1,\beta_k)},\bm\theta^{(s-1)},\phi^{(s-1)}\Big|\tilde{\bm{u}}\Big)\Big|_{\beta_k=\beta_k^{(s-1)}}d\\
& + \frac{1}{2} d^2 h_k^{(s)} +
  \lambda \|\bm{\beta}^{(s,s-1;k)} + d \bm{e}_{k}\|_1 \Big\}.
\end{align*}
\item [iii')] Choose a step size $\alpha_k^{(s)} >0$  and set
  $\bm{\beta}^{(s,s-1;k+1)} = \bm{\beta}^{(s,s-1;k)} + \alpha^{(s)}_k d_k^{(s)}
  \bm{e}_k$ such
  that $$Q_{\lambda}^{LA}\Big(\bm\beta^{(s,s-1;k+1)},\bm\theta^{(s-1)},\phi^{(s-1)}\big|\tilde{\bm{u}}\Big)
  \le Q_{\lambda}^{LA}\Big(\bm\beta^{(s,s-1;k)},\bm\theta^{(s-1)},\phi^{(s-1)}\big|\tilde{\bm{u}}\Big).$$
\end{itemize}
\end{algorithm}
We illustrate in the supplemental materials that the approximate GLMMLasso
algorithm speeds up remarkably without loosing that much
accuracy. Additionally, the approximation emphasizes the importance of a
refitting as advocated in the next section.

\subsection{Convergence behaviour and details of the GLMMLasso algorithm} 

\indent \textit{Numerical convergence.} The convergence of
the exact GLMMLasso algorithm to a stationary point can be proofed using
the results presented in \cite{Tseng09}. It is worth pointing out
that in the low-dimensional framework, the exact GLMMLasso algorithm with
$\lambda=0$ (no penalization) gives the same results as the function
\texttt{glmer} in the \texttt{R} package \texttt{lme4}.\\

\indent \textit{(0) Starting value $\bm\psi^{(0)}$.}
As starting value for $\bm\beta$, we fit a generalized
linear model with the Lasso where the regularization parameter is chosen by
cross-validation. The initial values for $\bm\theta$ and $\phi$ are then
calculated using steps (2) and (3) in Algorithm \ref{alg1} and
\ref{alg2}.\\ 

\indent \textit{i) Choice of $h_k^{(s)}$.}
For $h_k^{(s)}$ we choose the $k$th diagonal element of the Fisher
information of a generalized linear model. Hence we use the second
derivative of the first summand in (\ref{laplaceobfct}). We set $c_{min}
\le h_k^{(s)} \le c_{max}$ for positive constants $c_{min}$ and $c_{max}$
(e.g.\ $c_{min}=10^{-5}$ and $c_{max}=10^5$) in order that the algorithm
converges \citep{Tseng09}.\\

\indent \textit{ii) Calculation of $d_k^{(s)}$.}
The value $d_k^{(s)}$ is the minimizer of the quadratic approximation of the
objective function $Q_{\lambda}^{LA}(.)$ and analytically given by \citep{Tseng09}
\small
\begin{align}
d_k^{(s)}&= \begin{cases}
\displaystyle \median
 \Bigg(\frac{\lambda
   - \partial/\partial\beta_kf_{\beta_k}}{h_k^{(s)}},-\beta_{k},\frac{-\lambda-\partial/\partial\beta_k
   f_{\beta_k}}{h_k^{(s)}}\Bigg) & \text{if $\beta_k$ penalized}\\
-\frac{\partial/\partial_{\beta_k}f_{\beta_k}}{h_k^{(s)}}
& \text{otherwise},
\end{cases}
\end{align}
\normalsize
where $f_{\beta_k}=f\big(\bm\beta^{(s,s-1;k)},\bm\theta^{(s-1)},\phi^{(s-1)}\big)$
in Algorithm \ref{alg1} and
$f_{\beta_k}=f\big(\bm\beta^{(s,s-1;k)},\bm\theta^{(s-1)},\phi^{(s-1)}\big|
\tilde{\bm{u}}\big)$
in Algorithm \ref{alg2}.\\

\indent \textit{iii) Choice of $\alpha_k^{(s)}$.}  The step length
$\alpha_k^{(s)}$ is chosen such that the objective function
$Q_{\lambda}^{LA}(.)$ decreases. We suggest to use the Armijo rule, which
is defined for Algorithm \ref{alg1} as follows (and correspondingly for
Algorithm \ref{alg2} with fixed $\tilde{\bm{u}}$): \vspace{4mm}
\\
\textit{ Armijo rule: Choose $\alpha_k^{init} >0$ and let $\alpha_k^{(s)}$
  be the largest element of $\{\alpha_k^{init}\delta^l\}_{l=0,1,2,..}$
  satisfying
\begin{equation*}
Q_{\lambda}^{LA}\Big(\bm\beta^{(s,s-1;k)}+\alpha_k^{(s)}d_{k}^{(s)}\bm{e}_{k},\bm\theta^{(s-1)},\phi^{(s-1)}\Big)
\le
Q_{\lambda}^{LA}\Big(\bm\beta^{(s,s-1;k)},\bm\theta^{(s-1)},\phi^{(s-1)}\Big) +
\alpha_k^{(s)} \varrho \triangle^{k}
\end{equation*}
 where $\triangle^{k} := \partial/\partial\beta_k f_{\beta_k}d_k^{(s)} + \gamma (d_{k}^{(s)})^2
h_k^{(s)} + \lambda \|\bm{\beta}^{(s,s-1;k)}+d_{k}^{(s)}\bm{e}_{k}\|_1 - \lambda
\|\bm{\beta}^{(s,s-1;k)}\|_1$.
}
\vspace{2mm} \\
The choice of the constants comply with the suggestions in
\cite{BerD99}, e.g.\ $\alpha_k^{init}=1$, $\delta=0.5$, $\varrho=0.1$ and $\gamma=0$.\\

\indent\textit{Active Set Algorithm.}
If we assume that the true fixed-effect parameter $\bm\beta_0$ is sparse in
the sense that many elements are zero, we can reduce the computing
time remarkably by using an active set algorithm. This is also used in
\cite{Meie08} and \cite{Fried08}. In particular, we only cycle through all $p$
coordinates every $D$th iteration, otherwise only through the current
active set $S(\hat{\bm\beta}^{(s-1)})=\{k:\hat{\beta}_k^{(s-1)}\ne 0
\}$. Typical values for $D$ are $5$ and $10$.\\

\indent An implementation of the algorithm
is given in the \texttt{R} package \texttt{glmmixedlasso} and will be
made available on R-Forge (\texttt{http://r-forge.R-project.org/}).

\section{The two-stage GLMMLasso$^{LA}$ estimator(s)}  \label{Sec4} 
From the soft-thresholding property of the Lasso
in linear models \citep{Tibs96} and in Gaussian linear mixed models
\citep{Schell11}, the fixed-effect estimate $\hat{\bm\beta}$ is
biased towards zero. In some generalized linear mixed models the estimate
of the covariance parameter $\bm\theta$ is biased, too. To mitigate these
bias problems and the approximation error induced by using the approximate
GLMMLasso algorithm, we advocate a two-stage procedure. The first step aims at
estimating a candidate set of predictors $\hat{S}$ and can be seen as a variable screening procedure. The purpose of the second step is
a more unbiased estimation of the parameters using unpenalized maximum
likelihood (ML) estimation based on the selected variables  $\hat{S}$ from the
first step. The proposed two-stage GLMMLasso algorithm is summarized in Algorithm \ref{alg3}:\\
\begin{algorithm}
\caption{\textit{Two-stage GLMMLasso algorithm}}
\label{alg3}
\begin{itemize}
\item [] Stage 1: Compute the GLMMLasso$^{LA}$ estimate (\ref{laobfct}) and
  the set $\hat{S}$. 
\item [] Stage 2: Perform unpenalized ML estimation.
\end{itemize}
\end{algorithm}\\
\indent In the next subsections, we are going to discuss the specification of
the set of variables $\hat{S}$. We propose two methods from the
high-dimensional linear regression framework, and we do not consider the
adaptive Lasso \citep{Zou06}. 

\subsection{The GLMMLasso$^{LA}$-MLE hybrid estimator}
The LARS-OLS hybrid estimator was examined in \cite{Efro04} and also used in
\cite{Mein06} and \cite{Meie08}. In our context, it becomes a
two-stage procedure where the model is refitted including only the
covariates with a nonzero fixed-effect coefficient in
$\hat{\bm\beta}_{init}$, where
$(\hat{\bm\beta}_{init},\hat{\bm{\theta}}_{init},\hat{\phi}_{init})$ denotes
the initial estimate from (\ref{laobfct}). More specifically, choose
$\hat{S}=\hat{S}_{init}:=\{k : |\hat{\beta}_{k,init}\ne 0 \}$. Then the
GLMMLasso$^{LA}$-MLE hybrid estimator is given by
\begin{equation} \label{obfct2}
(\hat{\bm\beta},\hat{\bm{\theta}},\hat{\phi})_{hybrid} :=
\argmin_{\bm\beta_{\hat{S}_{init}},\bm\theta,\phi}  -2\log
L(\bm\beta_{\hat{S}_{init}},\bm\theta,\phi), 
\end{equation}
where for $S \subseteq \{1,\ldots,p\}$, $(\bm\beta_S)_k=\beta_k$ if $k \in S$
and $(\bm\beta_S)_k=0$ if $k \notin S$.

\subsection{The thresholded GLMMLasso$^{LA}$ estimator}
The thresholded Lasso with refitting in high-dimensional linear regression
models was examined in \cite{GeerBuhlZhou10} and \cite{Zhou10}.
We define the set $\hat{S}_{thres}$ to
be the set of variables which have initial fixed-effect coefficients larger than
some threshold $\lambda_{thres}>0$, i.e.\ we choose $\hat{S}=\hat{S}_{thres}:=\{k :
|\hat{\beta}_{k,init}| >\lambda_{thres}\}$. The thresholded
GLMMLasso$^{LA}$ estimator is then defined by
\begin{equation} \label{obfct3}
(\hat{\bm\beta},\hat{\bm{\theta}},\hat{\phi})_{thres} :=
\argmin_{\bm\beta_{\hat{S}_{thres}},\bm\theta,\phi} -2\log
L(\bm\beta_{\hat{S}_{thres}},\bm\theta,\phi).
\end{equation}
The thresholded GLMMLasso$^{LA}$ estimator involves another regularization
parameter $\lambda_{thres}$, which is determined by minimizing an
information criterion presented in the next subsection.

\subsection{Selection of the regularization parameters}
Estimators (\ref{laobfct}), (\ref{obfct2}) and (\ref{obfct3}) require the
choice of the regularization parameters $\lambda$ and $\lambda_{thres}$,
respectively. We propose to use the Bayesian Information Criterion (BIC) and
the Akaike Information Criterion (AIC), defined by
\begin{equation}
c_{n,\lambda} = -2\log L(\hat{\bm{\beta}},\hat{\bm{\theta}},\hat{\phi}) +
a(n) \cdot \hat{df}_{\lambda} 
\end{equation}
where $a(n)=\log(n)$ for the BIC and $a(n)=2$ for the AIC. Here,
$\hat{df}_{\lambda}=|\{1\le k \le p : \hat{\beta}_k \ne 0\}| +
\dim(\hat{\bm\theta})$ is the sum of of the number of nonzero fixed-effect
coefficients and the number of covariance parameters. The first summand is
motivated by the work of \cite{Zou07}. The second summand is the approach
of \cite{Bates10}, who proposes that in the classical generalized mixed
effects model the degrees of freedom are given by the number of unconstrained
optimization parameters. Based on our empirical experience, we suggest for
the estimators (\ref{laobfct}) and (\ref{obfct2}) the BIC, whereas for
(\ref{obfct3}) we advocate using the AIC (allowing for a larger number of
variables) to select $\lambda$ first and then, sequentially, the BIC to
select $\lambda_{thres}$. We will compare the performance of the three
estimators in the next sections.

\section{Simulation Study} \label{Sec5} In this section we assess the
performance of the GLMMLasso$^{LA}$ estimators (\ref{laobfct}),
(\ref{obfct2}) and (\ref{obfct3}). We compare them with appropriate Lasso,
maximum likelihood (ML) and Penalized Quasi-Likelihood
(PQL, \cite{Breslow93}) methods.\\
\indent In the main text, we only present simulation results for the
high-dimensional logistic mixed model. Simulation studies for the
low-dimensional logistic and the Poisson mixed model are included in the
supplementary material. At the end of this section, we compare the
GLMMLasso$^{LA}$ estimates in a situation where the number of noise
variables grows
successively.\\
\indent First of all, let us summarize some general conclusions drawn from
real data analysis and the simulation studies:
\begin{itemize}
\item [a)] The variable screening performance of the GLMMLasso algorithm is
  not only attractive for the high-dimensional setting, but also for
  low-dimensional data with a relatively large number of variables (say $p>20$).
\item [b)] The GLMMLasso algorithm is numerically as stable as standard
  \texttt{R} functions like \texttt{glmer} \citep{Bates10} or
  \texttt{glmmPQL} \citep{Breslow93,VenWR02} when $p<n$. On the other hand, 
  \texttt{glmpath} \citep{Park07} and \texttt{glmnet} \citep{Fried08} may
  fail to converge when high-dimensional models are misspecified.
\item [c)] The main difference between the logistic and the Poisson mixed
  model is the shrinkage of the covariance parameter estimates of the
  GLMMLasso$^{LA}$ estimator. These estimates are severely biased in
  logistic mixed models, in contrast to the Poisson mixed model. Further
  differences between these two classes are summarized in the supplemental
  materials.
\item [d)] The number of iterations $s$ substantially differs between the
  classes of generalized linear mixed models and the data set.
\end{itemize}

\subsection{Preview for the logistic mixed model}
In this section we confine the discussion to the logistic
mixed model because it is viewed as the most challenging model within the class
of generalized linear mixed models \citep{MolGV05,Jiang07}. As an overview,
let us sum up the main findings from the simulation study in the logistic
mixed model: 
\begin{itemize}
\item [i)] The GLMMLasso$^{LA}$ estimate from (\ref{laobfct}) of the
  covariance parameter $\bm\theta$ is notably biased. In other words,
  adding an $\ell_1$-penalty does not only shrink the fixed effects
  estimate $\hat{\bm\beta}$, but also the covariance parameter estimate
  $\hat{\bm\theta}$.
\item [ii)] In the high-dimensional settings, the GLMMLasso$^{LA}$-MLE
  hybrid estimator (\ref{obfct2}) performs better in terms of parameter
  estimation accuracy than the thresholded GLMMLasso$^{LA}$ estimator
  (\ref{obfct3}).
\item [iii)] The more random effects, the more important it is to use the
  GLMMLasso$^{LA}$ for variable screening (instead of a Lasso ignoring the
  grouping structure).
\item [iv)] The number of total iterations $s$ needed is small, often about
  15 iterations. 
\end{itemize}

\subsection{High-dimensional logistic mixed model}
In all subsequent simulation schemes (including the supplemental
materials), we restrict ourselves to the case where the number of
observations per cluster is equal, i.e.\ $n_r=n_C$ for $r=1,\ldots,N$. The
covariates are generated from a multivariate normal distribution with mean
zero and covariance matrix $\bm{V}$ with pairwise correlation
$\bm{V}_{kk'}=\rho^{|k-k'|}$ and $\rho=0.2$. Denote by $\bm\beta_0$ the
true fixed effects (wherein ($\bm\beta_0)_1$ is the intercept) and by $s_0$
the true number of
nonzero fixed-effect coefficients.\\
\indent For the logistic mixed models, the intercept and the first
covariate have independent random effects with different variance
parameters. In particular, $\bm\theta=(\theta_1,\theta_2)$ and covariance
matrix 
$\bm\Sigma_{\bm\theta}=\diag(\theta_1^2,,\ldots,\theta_1^2,\theta_2^2,\ldots,\theta_2^2) 
\in \mathbb{R}^{2N}$, i.e.\ $q=2N$. We investigate the following two
examples in the high-dimensional setting:
\begin{small}
\begin{itemize}
\item [$H_1$:] $N=40$, $n_C=10$, $n=400$, $p=500$,
  $\theta_1^2=\theta_2^2=1$ and $s_0=5$ with 
  $\bm\beta_0=(0.1,1,-1,1,-1,0,\ldots,0)^T$.
\item [$H_2$:] $N=50$, $n_C=10$, $n=500$, $p=1500$,
  $\theta_1^2=\theta_2^2=1$ and $s_0=5$ with 
  $\bm\beta_0=(0.1,1,-1,1,-1,0,\ldots,0)^T$.
\end{itemize}
\end{small}

\indent The fitted models are all correctly specified. Hereafter, we denote
by \textit{oracle} the ML estimate of the model which includes only the
variables from the true active set. Let \textit{glmmlasso}, \textit{hybrid
  glmmlasso} and \textit{thres glmmlasso} be the GLMMLasso$^{LA}$ estimates
(\ref{laobfct}), (\ref{obfct2}) and (\ref{obfct3}), respectively. We
compare the GLMMLasso$^{LA}$ methods with the standard Lasso for
generalized linear models (which ignore the grouping structure). For that
purpose we use the \textit{glmpath} algorithm \citep{Park07} and the BIC as
variable selection criterion. Then, let \textit{hybrid glmpath} and
\textit{thres glmpath} be the two-stage
procedures based on \textit{glmpath} (without random effects).\\
\indent The results in the form of median and rescaled median absolute
deviation (in 
parentheses) over 100 simulation runs are shown in Table
\ref{table2}. There, $|S(\hat{\bm\beta})|$ denotes the cardinality of the
estimated active set and TP is the number of true positives (selected
variables which are in the true active set). SE is the squared error
of the fixed-effect coefficients, i.e.\ SE$=\|\hat{\bm \beta} -
\bm\beta_0\|_2^2$.
\begin{table}[!h]
\footnotesize
\begin{center}
  \caption{\textit{Simulation results (medians) for the logistic mixed
      models $H_1$ and $H_2$ (rescaled median absolute deviations in
      parentheses).
      A $^*$ means that the corresponding coefficient is not subject to
      penalization in the GLMMLasso$^{LA}$ estimate.}} \label{table2}
  \vspace{0.2cm}
\begin{tabular}{clcccccccccc}
\hline \hline
Model & Method & $|S(\hat{\bm\beta})|$ & TP & $\hat{\theta}_1^2$
& $\hat{\theta}_2^2$ & $\hat{\beta}_{1}^*$ & $\hat{\beta}_{2}^*$ &
$\hat{\beta}_{3}$ & 
$\hat{\beta}_{4}$ & $\hat{\beta}_{5}$ & SE \\ \hline \vspace{3mm}
True&& 5 & 5 & 1 & 1 & 0.1 & 1 & -1 & 1 & -1 \\
$H_1$ & oracle & 5 & 5 & 0.85 & 0.86 & 0.07 & 1.04 & -0.99 & 0.98 & -1.01 & 0.14 \\ 
   &  & (0) & (0) & (0.4) & (0.59) & (0.2) & (0.25) & (0.22) & (0.18) & (0.14) & (0.088) \\ 
   & glmmlasso & 6 & 5 & 0.38 & 0.37 & 0.06 & 0.66 & -0.3 & 0.26 & -0.34 & 1.6 \\ 
   &  & (1.48) & (0) & (0.24) & (0.3) & (0.14) & (0.16) & (0.14) & (0.14) & (0.12) & (0.42) \\ 
   & glmpath & 7 & 5 & - & - & 0.04 & 0.24 & -0.21 & 0.22 & -0.28 & 2.4 \\ 
   &  & (2.22) & (0) & - & - & (0.13) & (0.12) & (0.11) & (0.1) & (0.1) & (0.52) \\ 
   & hybrid glmmlasso & 6 & 5 & 0.89 & 0.87 & 0.08 & 1.05 & -0.99 & 1 & -1.03 & 0.44 \\ 
   &  & (1.48) & (0) & (0.43) & (0.58) & (0.19) & (0.25) & (0.23) & (0.18) & (0.16) & (0.32) \\ 
   & hybrid glmpath & 7 & 5 & 0.86 & 0.87 & 0.08 & 1.01 & -0.99 & 0.99 & -1.02 & 0.7 \\ 
   &  & (2.22) & (0) & (0.42) & (0.53) & (0.2) & (0.28) & (0.24) & (0.19) & (0.16) & (0.64) \\ 
   & thres glmmlasso & 10 & 5 & 1.02 & 1.11 & 0.1 & 1.19 & -1.09 & 1.11 & -1.13 & 1.3 \\ 
   &  & (3.71) & (0) & (0.7) & (0.85) & (0.22) & (0.29) & (0.23) & (0.2) & (0.19) & (0.77) \\ 
   & thres glmpath & 10 & 5 & 0.91 & 0.94 & 0.09 & 1.11 & -1.07 & 1.11 & -1.1 & 1.1 \\ \vspace{3mm}
   &  & (2.97) & (0) & (0.49) & (0.59) & (0.21) & (0.27) & (0.25) & (0.19)
   & (0.2) & (0.73) \\ 

$H_2$ & oracle & 5 & 5 & 0.89 & 0.94 & 0.11 & 1.02 & -0.98 & 1.02 & -1.02 & 0.13 \\ 
   &  & (0) & (0) & (0.4) & (0.53) & (0.18) & (0.25) & (0.15) & (0.18) & (0.16) & (0.1) \\ 
   & glmmlasso & 6 & 5 & 0.39 & 0.41 & 0.09 & 0.66 & -0.31 & 0.27 & -0.34 & 1.6 \\ 
   &  & (1.48) & (0) & (0.23) & (0.28) & (0.13) & (0.17) & (0.1) & (0.11) & (0.09) & (0.27) \\ 
   & glmpath & 6.5 & 5 & - & -  & 0.08 & 0.23 & -0.21 & 0.21 & -0.28 & 2.4 \\ 
   &  & (0.74) & (0) & - & - & (0.11) & (0.13) & (0.08) & (0.11) & (0.08) & (0.34) \\ 
   & hybrid glmmlasso & 6 & 5 & 0.93 & 0.96 & 0.12 & 1.02 & -0.99 & 1.05 & -1.04 & 0.34 \\ 
   &  & (1.48) & (0) & (0.44) & (0.51) & (0.19) & (0.26) & (0.15) & (0.17) & (0.16) & (0.3) \\ 
   & hybrid glmpath & 6.5 & 5 & 0.87 & 0.94 & 0.12 & 1.01 & -0.99 & 1.03 & -1.04 & 0.48 \\ 
   &  & (0.74) & (0) & (0.42) & (0.5) & (0.18) & (0.22) & (0.15) & (0.18) & (0.17) & (0.37) \\ 
   & thres glmmlasso & 14 & 5 & 1.3 & 1.33 & 0.16 & 1.26 & -1.16 & 1.2 & -1.22 & 2 \\ 
   &  & (5.93) & (0) & (0.87) & (0.79) & (0.27) & (0.27) & (0.28) & (0.26) & (0.24) & (1.7) \\ 
   & thres glmpath & 13.5 & 5 & 0.9 & 1.03 & 0.17 & 1.17 & -1.07 & 1.13 & -1.15 & 1.8 \\ 
   &  & (5.19) & (0) & (0.52) & (0.64) & (0.24) & (0.25) & (0.19) & (0.22) & (0.21) & (1.2) \\ \hline 
\end{tabular}
\end{center}
\end{table}

Comparing the cardinality of the active set, we see that \textit{thres
  glmmlasso} and \textit{thres glmpath} have much larger active sets than
\textit{glmmlasso} and \textit{glmpath}, respectively. This is largely due
to the fact that we employ the AIC in the first and the BIC in the second
stage. This is outweighed by the advantage that on average (not shown), the
true effects are predominantly included in \textit{thres
  glmmlasso}. 
The active set of \textit{glmmlasso} is slightly smaller than that of
\textit{glmpath}. And yet, the number of TP is similar as for
\textit{glmpath}. Hence, we conclude that the existence of random
effects does affect the variable selection performance of \textit{glmpath}.\\
\indent Concerning covariance parameter estimation, we read off from the
table that $\hat{\theta}_1^2$ and $\hat{\theta}_2^2$ are seriously biased
for \textit{glmmlasso}. This motivates the usage of a two-stage
procedure. The table suggests that the hybrid and the thresholded
procedures have improved estimation accuracy of the random effects
parameters compared to their original counterparts.\\ 
\indent Looking at the fixed-effect parameter estimation accuracy, the
simulation study reveals that the \textit{glmmlasso} estimates are less
biased than the corresponding \textit{glmpath} estimates, resulting in
lower squared error. And the same holds for \textit{hybrid glmmlasso}
and \textit{hybrid glmpath}. The fixed-effect parameter estimates of
\textit{thres glmmlasso} and \textit{thres glmpath} perform inadequately
compared to their \textit{hybrid} counterparts. As marked by an asterisk in
the table, $\beta_2$ is not subject to penalization for the
GLMMLasso$^{LA}$ estimator since this variable has a random effect
\citep{Schell11}. Thus the bias of the
estimate is much smaller than for the other fixed-effect coefficients.\\
\indent To sum up the simulation study, we first conclude that \textit{hybrid
  glmmlasso} outperforms \textit{thres glmmlasso} in terms of parameter
estimation accuracy, with similar performance regarding true
positives. Second, \textit{glmmlasso} procedures do outperform
\textit{glmpath} procedures as variable screening methods. 
Of course, \textit{glmpath} is fitting a wrong model without random effects.

\subsection{Logistic mixed model with a growing number of noise covariates}  
Here, we assess the performance of \textit{glmmlasso} and \textit{hybrid
  glmmlasso} when the number of noise variables grows successively. In the
low-dimensional setting, we compare them with the ML estimate computed by
the \texttt{R} function \texttt{glmer} (denoted by \textit{glmer}). In
addition, let \textit{p-glmer} be the method which performs variable
selection in the following way: Eliminate consecutively (backward
selection) all variables with a p-value larger than $5\%$ until the final
model is attained comprising only significant variables. We compare these
four methods in terms of their performance of twice the negative
out-of-sample log-likelihood. Let us fix the following random intercept
model design: $n=400$, $N=40$, $n_C=10$, $\theta^2=1$,
$\bm\beta_0=(0,1,-1,1,-1)$. We start with $p=5$ (no noise variables) and
raise the number of variables to $p=65$. The results over 50 simulation
runs are depicted in Figure 1.

\begin{figure}
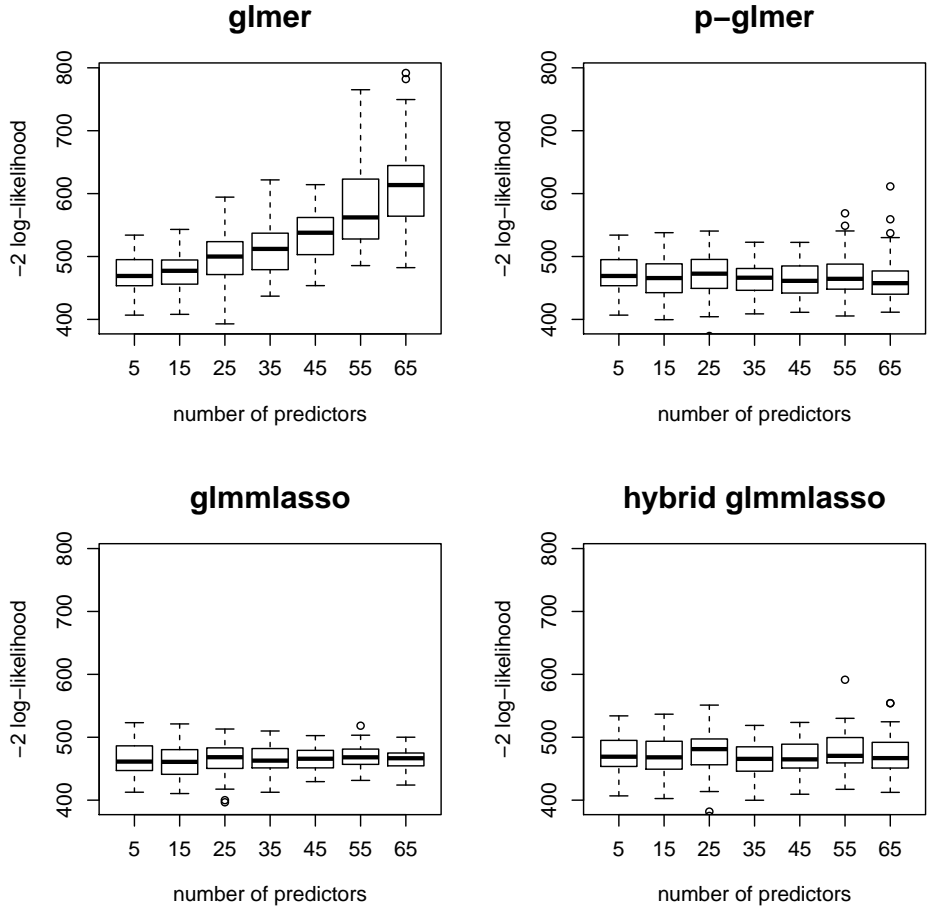

\epsfCfile{0.8}{growingP}
\caption{Minus twice out-of-sample log-likelihood for a growing number of
  covariates. The ML estimate performs badly whereas the GLMMLasso$^{LA}$
  estimators remain stable, and they are comparable to the p-glmer in the
  low-dimensional framework.}
\end{figure}

The figures show that the negative out-of-sample log-likelihood values for
\textit{glmer} grow polynomial whereas the likelihoods for the other
methods remain fairly constant. The increase in \textit{glmer} stems from
the fact that it overfits the model for a growing number of covariates.
When focusing on the figures in more detail, we read off that the negative
log-likelihood of \textit{glmmlasso} increases slightly for larger $p$
whereas the negative log-likelihood of \textit{hybrid glmmlasso} remains
stable. The rationale for this small increase in \textit{glmmlasso} is that
the more noise covariates, the larger the optimal $\lambda$, and henceforth
the larger the shrinkage of the fixed effects. And this leads to the
increase of the out-of-sample log-likelihood. \textit{hybrid glmmlasso}
(and also \textit{thres glmmlasso}) overcomes this problem and leads to a
stable out-of-sample log-likelihood irrespective of $p$.

\subsection{Correlated Random Effects}
Both from a methodological and an implementational point of view it is
conceptually possible to use correlated random effects. As an
illustration we use the logistic mixed model $H_1$ with correlated random
effects (with unstructured covariance matrix) where we use a correlation of
$\rho = 0.5$ between the two random effects. The corresponding results are
illustrated in Table \ref{table2cor}. The results are very similar to the
uncorrelated case. However, the bias of the correlation estimate seems to
be less severe than the bias of the variance components.
\begin{table}[!h]
\footnotesize
\begin{center}
  \caption{\textit{Simulation results (medians) for the logistic mixed
      models $H_1$ (rescaled median absolute deviations in parentheses).
      A $^*$ means that the corresponding coefficient is not subject to
      penalization in the GLMMLasso$^{LA}$ estimate.}} \label{table2cor}
  \vspace{0.2cm}
\begin{tabular}{clccccccccccc}
\hline \hline
Model & Method & $|S(\hat{\bm\beta})|$ & TP & $\hat{\theta}_1^2$
& $\hat{\theta}_2^2$ & $\hat{\rho}$ & $\hat{\beta}_{1}^*$ & $\hat{\beta}_{2}^*$ &
$\hat{\beta}_{3}$ & $\hat{\beta}_{4}$ & $\hat{\beta}_{5}$ & SE \\ \hline
\vspace{3mm} 
True&& 5 & 5 & 1 & 1 & 0.5 & 0.1 & 1 & -1 & 1 & -1 \\

$H_1$ & oracle & 5 & 5 & 0.88 & 0.94 & 0.53 & 0.1 & 0.97 & -1.03 & 1.02 & -1.01 & 0.14 \\ 
   &  & (0) & (0) & (0.46) & (0.54) & (0.37) & (0.18) & (0.24) & (0.17) & (0.15) & (0.15) & (0.1) \\ 
   & glmmlasso & 6 & 5 & 0.41 & 0.41 & 0.63 & 0.07 & 0.66 & -0.33 & 0.28 & -0.34 & 1.6 \\ 
   &  & (1.48) & (0) & (0.22) & (0.25) & (0.51) & (0.14) & (0.16) & (0.12) & (0.11) & (0.11) & (0.35) \\
   \hline
\end{tabular}
\end{center}
\end{table}

\section{Illustration}  \label{Sec6}
In this section we illustrate the proposed GLMMLasso$^{LA}$ estimators for
Poisson regression on an extended real data set with count data.\\

\textit{Data description.} We consider
the epilepsy data from \cite{Thall90} which were also analyzed by
\cite{Breslow93}. The data were obtained from a randomized clinical trial of 59
patients with epilepsy, comparing a new drug (Trt=1) with placebo
(Trt=0). The response variable consists of counts of epileptic seizures
during the two weeks before each of four clinic visits (V4=1 for fourth
visit, 0 otherwise). Further covariates in the analysis are the logarithm
of age (Age), the logarithm of 1/4 the number of baseline seizures (Base)
and the interaction of Base and Trt (Base x Trt). The main question of
interest is whether taking the new drug reduces the number of epileptic
seizures compared with placebo. In order to assess the performance of the
proposed procedure with high-dimensional data, we add $U(-1,1)$ distributed
noise predictors to get a data set with $n=236$, $N=59$, $n_r=4$ for
$r=1,\ldots,N$ and $p=4000$. All predictors are standardized to have mean
zero and standard deviation one.\\ 

\textit{Model.} Model III in \cite{Breslow93} is a two level GLMM
\citep{Bates10}, which is an extension of the single level GLMM introduced
in Section 2 for more than one grouping variable. The model consists of two
independent random intercept effects. One for subject (level 1, index $r$)
and one for observation (level 2, index
$j$).  Let $\theta_{sub}^2$
and $\theta_{obs}^2$ be the corresponding variance parameters.
 Then the linear
predictor can be written as $$\log(\mu_{rj}) =\eta_{rj}= \bm{x}_{rj}^T\bm\beta +
\theta_{sub} u_r + \theta_{obs} u_{rj} \quad r=1,\ldots,59,\quad j=1,\ldots,4.$$

\textit{Results.} The results of the analysis are presented in Table
\ref{tableAppl}. In the first column we show the estimates for Model III
without performing variable selection. There, Intercept, Base and Trt are
significant at the $5\%$ level (indicated by $^{\dagger}$). If we perform
backward selection using the BIC, we end up with a model including Intercept
and Base only. And this model coincides with the one selected by
\textit{glmmlasso}. \textit{Hybrid glmmlasso} overcomes the bias
problems of \textit{glmmlasso} and it yields a better model in terms of the
BIC. \textit{Thres glmmlasso} includes additional noise variables, thereby achieving the smallest BIC score for all models
under consideration. Comparing \textit{hybrid glmmlasso} and
\textit{thres glmmlasso}, the table suggests that the additional covariates
in the latter model reduce the variability while keeping the fixed-effect
estimates unaltered.
\begin{table}[!h]
\footnotesize
\begin{center}
\caption{\textit{Results for the epilepsy data. Model III is based on 6
fixed-effect covariates while the other methods are based on $p=4000$
variables, including 3994 noise covariates. $^{\dagger}$ indicates that the
    corresponding coefficient is significant at the $5\%$
    level. $^{\ddagger}$ means that five noise variables are selected,
    but not shown in the table. $S(\hat{\bm\beta})=\{k:\hat{\beta}_k \ne 0
\}$ is the total number of selected variables.}} \label{tableAppl}
\vspace{0.2cm}
\begin{tabular}{lcccc} \hline \hline
 & Model III & glmmlasso & hybrid glmmlasso & thres glmmlasso\\
\hline
BIC &527.3 & 571.8 & 515.5 & 480.3\\
$S(\hat{\bm\beta})$&6&2&2&$7^{\ddagger}$ \\
Intercept&$1.58^{\dagger}$&1.62&1.58&1.58 \\
Base&$0.66^{\dagger}$&$<10^{-4}$&0.74&0.75 \\
Trt&$-0.47^{\dagger}$&-&-&- \\
Base x Trt&0.36&-&-&- \\
Age &0.11&-&-&- \\
V4 &-0.04&-&-&- \\
$\hat{\theta}_{sub}^2$ &0.21&0.68&0.25&0.28 \\
$\hat{\theta}_{obs}^2$ &0.13&0.12&0.13&0.04 \\
\hline
\end{tabular}
\end{center}
\end{table}

\section{Concluding Remarks} \label{Sec7} We address the problem of
estimating high-dimensional generalized linear mixed models (GLMMs). While
low-dimensional generalized linear mixed models \citep{Bates10} and
high-dimensional generalized linear models \citep{Geer08} have been
extensively studied in recent years, little attention has been devoted to
high-dimensional GLMMs. We provide an efficient algorithm for the
$\ell_1$-penalized maximum likelihood estimator, called GLMMLasso. It is
based on the Laplace approximation, coordinatewise optimization and a
speeding up approximation. The method should be typically used as a
screening procedure to estimate a small set of important variables. We
propose refitting by maximum likelihood to get accurate parameter
estimates. The second stage is much more important than for linear models,
because $\ell_1$-shrinkage can lead to severe bias problems for the
estimation of the variance components. Our work is primarily a contribution
addressing the numerical challenges of performing high-dimensional variable
selection and parameter estimation in nonlinear mixed-effects models
involving a non-convex loss function. An implementation of the algorithm
can be found in our \texttt{R} package \texttt{glmmixedlasso}. It will be
made available on R-Forge.



\bigskip
\begin{center}
  {\large\bf ACKNOWLEDGEMENTS}
\end{center}
The research is supported in part by the Swiss National Science Foundation
(grant no. 20PA21-120043/1, ``Forschergruppe FOR 916''). The authors thank
the members of the DFG-SNF Forschergruppe 916 for many stimulating
discussions. In particular we would like to thank Stephan Dlugosz from the
ZEW Mannheim for insisting on studying this particular kind of
problem. Moreover, the authors thank the associate editor and two referees
for their helpful comments.

\bibliography{glmmLit}

\newpage
\begin{center}
\emph{Appendices to} \\\vspace{0.5cm}

\Large{\textbf{"GLMMLasso: An Algorithm for High-Dimensional
  Generalized Linear Mixed Models Using $\ell_1$-Penalization"}}\\\vspace{0.5cm}
 
\normalsize
J\"urg Schelldorfer, Lukas Meier and Peter B\"uhlmann
\end{center}

\setcounter{table}{3}
\section*{Appendix A: PIRLS algorithm}
In this section, we explain how to determine
the mode $\tilde{\bm{u}}=\argmax_{\bm{u}}- S(\bm{u})$ (introduced in
Section 2 of the article). We have to solve the following
minimization problem:
\begin{equation} \label{120510}
\tilde{\bm{u}} = \argmin_{\bm{u}} S(\bm{u}) := - \sum_{i=1}^n \Bigg\{ \frac{y_{i}\xi_i(\bm{u}) - b(\xi_i(\bm{u}))
}{\phi} + c(y_i,\phi) \Bigg\} + \frac{1}{2}\|\bm{u}\|_2^2.
\end{equation}
We would like to highlight that $S(\bm{u})$ is a convex function. 
We employ the Newton-Raphson algorithm to find a global minimum. From
(\ref{120510}) we get 
\begin{equation*}
S'(\bm{u})= -(\bm{Z}\bm\Lambda_{\bm\theta})^T\bm{B}(\bm{y}-\bm{\mu}) +\bm{u} \quad
, \quad S''(\bm{u}) = (\bm{Z}\bm\Lambda_{\bm\theta})^T \bm{W}
(\bm{Z}\bm\Lambda_{\bm\theta}) + \bm{1}_q
\end{equation*}
where $\bm{W}=\diag^{-1}\Big(\phi v(\mu_i)g'(\mu_i)^2\Big)_{i=1}^n$,
$\bm{B}=\diag^{-1}\Big(\phi v(\mu_i)g'(\mu_i)\Big)_{i=1}^n$ and  $v(.)$ is
the conditional variance function \citep{McCPN89}. Then following the lines
in \citet{HasTTF09}, we get Algorithm \ref{alg4}, which is also described
in \cite{Bates09c} and \cite{Bates09b}. 

\setcounter{algorithm}{3}
\begin{algorithm}
\caption{\textit{PIRLS algorithm}}
\label{alg4}
Choose a starting value or set $\bm{u}^{(0)}=0$.\\
\textbf{Repeat} for $r=0,1,2,\ldots$
\begin{flalign*}
& \bm\eta^{(r)} = \bm{X}\bm\beta +
\bm{Z}\bm\Lambda_{\bm\theta}\bm{u}^{(r)}\\
&\bm\mu^{(r)} = g^{-1}(\bm\eta^{(r)}) \\
&\bm{W}^{(r)} = \diag \Bigg(\frac{1}{\phi v(\mu_i^{(r)})g'(\mu_i^{(r)})^2}\Bigg)_{i=1}^n \quad
\bm{G}^{(r)}=\diag\Big(\phi v(\mu_i^{(r)})g'(\mu_i^{(r)})^2\Big)_{i=1}^n\\
& \bm{B}^{(r)} = \diag \Bigg(\frac{1}{\phi v(\mu_i^{(r)})g'(\mu_i^{(r)})}\Bigg)_{i=1}^n\\
&\bm{z}^{(r)} = (\bm{Z}\bm\Lambda_{\bm\theta})\bm{u}^{(r)} +
\bm{G}^{(r)}\bm{B}^{(r)} (\bm{y}-\bm\mu^{(r)})
\end{flalign*}
\quad \quad Then solve
\begin{equation*}
\Big((\bm{Z}\bm\Lambda_{\bm\theta})^T\bm{W}^{(r)}(\bm{Z}\bm\Lambda_{\bm\theta}) + \bm{1}_q
\Big)\bm{u}^{(r+1)}=(\bm{Z}\bm\Lambda_{\bm\theta})^T \bm{W}^{(r)} \bm{z}^{(r)}
\end{equation*}
\textbf{until} $$\frac{\|\bm\eta^{(r+1)}-\bm\eta^{(r)}\|_2}{\|\bm\eta^{(r)}\|_2}
\le tol \quad .$$\\
Set $\tilde{\bm{u}}=\bm{u}^{(r+1)}$.\\
\end{algorithm}
The PIRLS algorithm typically converges fast. To further speed up Algorithm
1 and 2, we use the current value of $\tilde{\bm{u}}$ as starting value in
step (1) a) of Algorithm 1. Consequently, the number of iterations required
to update $\tilde{\bm{u}}$ is indeed small, often smaller than three.

\section*{Appendix B: Comparison of the exact and approximate GLMMLasso algorithm}
In this section, we compare the exact and the
approximate algorithm (i.e.\ Algorithm 1 and 2) on various simulated data
sets. We use the same model settings as in the simulation studies (see
Section 5 of the main article and Appendices C and D).\\
\indent First of all, let us give an overview of the key findings about the
approximate version of the algorithm:
\begin{enumerate}
\item The approximate algorithm is substantially faster than the exact
  algorithm (often more than 50\%).
\item For the logistic mixed model, the loss in accuracy (with respect to
  variable selection and parameter estimation) is very small.
\item For the Poisson mixed model, the loss in accuracy stems from the
  selection of too many covariates with very small fixed-effects
  coefficients. This problem is effectively alleviated by the proposed
  two-stage procedures (see Section 4).
\end{enumerate}

\indent In detail, we compare the
algorithms in terms of computing time, number of iterations, likelihood
function, the active set and the fixed-effects estimation accuracy. Denote
by $x^e$ the measure for the exact and $x^a$ the corresponding measure for
the approximate GLMMLasso algorithm. Precisely, let 
$rel.time=t^{a}/t^{e}$ be the relative (cpu) time,
$rel.iter=Iter^{a}/Iter^{e}$ be the relative number of outer
iterations s, $rel.ll=|\ell^a-\ell^e|/|\ell^e|$ the relative difference of
the likelihood function values, $rel.fix=\|\bm\beta^{a}-\bm\beta^{e} \|_2/\|\bm\beta^e
\|_2$ be the relative difference of the fixed-effects parameters and
$activeSet$ the percentage of models where the active sets completely
coincide for the exact and the approximate algorithm. For each model, we
carry out 50 simulation runs. And for each run, we compare the results of
the algorithm on a sequence of 21 $\lambda$-values. The results in the form
of means and standard deviations (in parentheses) are depicted in Table
\ref{table10} (logistic mixed model) and Table \ref{table11} (Poisson mixed model).
\begin{table}[!h]
\begin{center}
\caption{\textit{Simulation results (mean values, standard deviations in
    parentheses) for logistic mixed models (Section
    5 and Appendix C).}} \label{table10}
\vspace{0.2cm}
\begin{tabular}{cccccc}
\hline \hline
Model & $rel.time$& $rel.iter$ & $rel.ll$ & $rel.fix$ & $activeSet$ \\ 
\hline
L1 &0.58&0.63&$6 \times 10^{-4}$&0.02&0.98 \\
&(0.18)&(0.28)&($3 \times 10^{-4}$)&(0.01)&(0.04) \\
L2 &0.41&0.61&$8 \times 10^{-4}$&0.02&0.87 \\
&(0.09)&(0.18)&($5 \times 10^{-4}$)&(0.01)&(0.07) \\
H1 &0.21&0.67&$9 \times 10^{-4}$&0.02&0.83 \\
&(0.07)&(0.32)&($7 \times 10^{-4}$)&(0.01)&(0.09) \\
H2 &0.28&0.77&$8 \times 10^{-4}$&0.02&0.84 \\
&(0.14)&(0.96)&($6 \times 10^{-4}$)&(0.01)&(0.08) \\
\hline
\end{tabular}
\end{center}
\end{table}

\begin{table}[!h]
\begin{center}
\caption{\textit{Simulation results (mean values, standard deviations in
    parentheses) for Poisson mixed models (Appendix D).}} \label{table11}
\vspace{0.2cm}
\begin{tabular}{cccccc}
\hline \hline
Model & $rel.time$& $rel.iter$ & $rel.ll$ & $rel.fix$ & $activeSet$ \\ 
\hline
L1 &0.10&0.10&$46 \times 10^{-4}$&0.12&0.96 \\
&(0.02)&(0.02)&($91 \times 10^{-4}$)&(0.07)&(0.11) \\
L2 &0.06&0.11&$27 \times 10^{-4}$&0.13&0.88 \\
&(0.01)&0.02&($8 \times 10^{-4}$)&(0.10) &(0.10) \\
H1 &0.11&0.17&$524 \times 10^{-4}$&0.33 &0.30 \\
&(0.02)&(0.03)&($697 \times 10^{-4}$)&(0.35)&(0.26) \\
H2 &0.09&0.17&$698 \times 10^{-4}$&0.38&0.31 \\
&(0.02)&0.04&($951 \times 10^{-4}$)&(0.38)&(0.29) \\
H3 &0.19&0.92&$1296 \times 10^{-4}$&0.10&0.05 \\
&(0.07)&(0.39)&($451 \times 10^{-4}$)&(0.03)&(0.11) \\
\hline
\end{tabular}
\end{center}
\end{table}
We see for both the logistic and the Poisson mixed model that the approximate
algorithm requires noteworthy less computing time and outer iterations. The
gain in computing time is impressive and often more than 50\%. It is
apparent that the two procedures yield similar likelihood function values,
although the Poisson mixed model has larger differences than the logistic mixed model. We read off from Table \ref{table10} that the
parameter estimates are very similar and that the active sets coincide
well. Table \ref{table11} suggests that the active sets and the
parameter estimates differ considerably more between the exact and the
approximate algorithm. By a closer look, we do see that the differences
originate in the fact that for some data sets the approximate algorithm
selects more variables, but with very small fixed-effects
coefficients. This explains the low values of \textit{activeSet}. This
problem can be effectively addressed by the two-stage procedures presented
in Section 4 of the article.\\
\indent To sum
up, the simulations do not only encourage the attractiveness of the approximate
algorithm with respect to speed, but also the need for the two-stage procedures.   

\section*{Appendix C: Low-dimensional logistic mixed model}
In the low-dimensional setting, we compare our methods with the unpenalized
maximum likelihood (ML) estimate and the Penalized Quasi-Likelihood (PQL,
\cite{Breslow93}) estimate. We denote them by \textit{glmer} and
\textit{glmmPQL}, respectively. The comparison begs the question of how to
perform variable selection for \textit{glmer} and \textit{glmmPQL}.  We
need some kind of variable selection procedure such that the results remain
comparable with our methods. Hence we suggest to reduce iteratively
(backward selection) the number of covariates by dropping those whose
p-value is greater than $5\%$. By doing so, we end up with a model where
all variables are significant. We denote these methods by \textit{p-glmer}
and
\textit{p-glmmPQL}.\\
We present the following two examples in the low-dimensional setting:
\begin{itemize}
\item [$L_1$:] $N=40$, $n_C=10$, $n=400$, $p=10$, $\theta_1^2=\theta_2^2=1$
  and $s_0=5$ with 
  $\bm\beta_0=(0.1,1,-1,1,-1,0,\ldots,0)^T$.
\item [$L_2$:] $N=40$, $n_C=10$, $n=400$, $p=50$, $\theta_1^2=\theta_2^2=1$
  and $s_0=5$ with 
  $\bm\beta_0=(0.1,1,-1,1,-1,0,\ldots,0)^T$.
\end{itemize}
The results in the form of median and rescaled median absolute deviation (in
parentheses) over 100 simulation runs are depicted in Table
\ref{table12}. There, $|S(\hat{\bm\beta})|$ denotes the cardinality of the
estimated active set and TP is the number of true positives (selected
variables which are in the true active set). SE is the squared error
of the fixed-effect coefficients.

\begin{table}[!h]
\footnotesize
\begin{center}
\caption{\textit{Simulation results for the logistic mixed models $L_1$ and
    $L_2$ (rescaled median absolute deviations in parentheses). A $^*$ means that the
    corresponding coefficient is not subject to $\ell_1$-penalization in
    the GLMMLasso$^{LA}$ estimate.}} \label{table12}
\vspace{0.2cm}
\begin{tabular}{clcccccccccc}
\hline \hline
Model & Method& $|S(\hat{\bm\beta})|$ & TP & $\hat{\theta}_1^2$
& $\hat{\theta}_2^2$ & $\hat{\beta}_{1}^*$ & $\hat{\beta}_{2}^*$ &  $\hat{\beta}_{3}$ &
$\hat{\beta}_{4}$ & $\hat{\beta}_{5}$ & SE \\ \hline \vspace{3mm}
True&& 5 & 5 & 1 & 1 & 0.1 & 1 & -1 & 1 & -1 & \\  

$L_1$ & oracle & 5 & 5 & 0.92 & 0.82 & 0.09 & 0.99 & -1 & 0.99 & -0.98 & 0.18 \\ 
   &  & (0) & (0) & (0.41) & (0.43) & (0.18) & (0.32) & (0.19) & (0.19) & (0.16) & (0.16) \\ 
   & glmmlasso & 6 & 5 & 0.65 & 0.54 & 0.08 & 0.82 & -0.72 & 0.68 & -0.74 & 0.4 \\ 
   &  & (1.48) & (0) & (0.35) & (0.33) & (0.15) & (0.24) & (0.17) & (0.17) & (0.16) & (0.23) \\ 
   & glmpath & 6 & 5 & - & - & 0.05 & 0.51 & -0.54 & 0.53 & -0.57 & 0.88 \\ 
   &  & (1.48) & (0) & - & - & (0.13) & (0.16) & (0.17) & (0.15) & (0.15) & (0.46) \\ 
   & hybrid glmmlasso & 6 & 5 & 0.97 & 0.83 & 0.08 & 1.02 & -1.01 & 0.99 & -1 & 0.28 \\ 
   &  & (1.48) & (0) & (0.43) & (0.47) & (0.17) & (0.31) & (0.19) & (0.2) & (0.18) & (0.21) \\ 
   & hybrid glmpath & 6 & 5 & 0.95 & 0.81 & 0.08 & 1 & -1.01 & 1 & -1 & 0.25 \\ 
   &  & (1.48) & (0) & (0.43) & (0.49) & (0.18) & (0.31) & (0.19) & (0.19) & (0.17) & (0.2) \\ 
   & thres glmmlasso & 5 & 5 & 0.96 & 0.83 & 0.08 & 0.99 & -1 & 0.99 & -0.98 & 0.2 \\ 
   &  & (0) & (0) & (0.42) & (0.44) & (0.18) & (0.32) & (0.18) & (0.19) & (0.16) & (0.18) \\ 
   & thres glmpath & 5 & 5 & 0.92 & 0.83 & 0.08 & 0.99 & -1 & 0.99 & -0.98 & 0.2 \\ 
   &  & (0) & (0) & (0.41) & (0.44) & (0.18) & (0.32) & (0.19) & (0.19) & (0.16) & (0.18) \\ 
   & p-glmer & 5 & 5 & 0.94 & 0.83 & 0.08 & 1 & -1.01 & 0.99 & -0.98 & 0.23 \\ 
   &  & (0) & (0) & (0.42) & (0.48) & (0.18) & (0.31) & (0.2) & (0.19) & (0.16) & (0.2) \\ 
   & p-glmmPQL & 5 & 5 & 0.97 & 0.87 & 0.08 & 0.9 & -0.94 & 0.93 & -0.92 & 0.19 \\ \vspace{3mm}
   &  & (0) & (0) & (0.42) & (0.4) & (0.17) & (0.27) & (0.19) & (0.18) & (0.14) & (0.16) \\

$L_2$ & oracle & 5 & 5 & 0.95 & 0.8 & 0.07 & 0.99 & -0.98 & 0.99 & -1.03 & 0.16 \\ 
   &  & (0) & (0) & (0.47) & (0.39) & (0.22) & (0.29) & (0.18) & (0.2) & (0.16) & (0.1) \\ 
   & glmmlasso & 6 & 5 & 0.47 & 0.36 & 0.05 & 0.71 & -0.46 & 0.47 & -0.49 & 0.99 \\ 
   &  & (1.48) & (0) & (0.26) & (0.28) & (0.17) & (0.2) & (0.13) & (0.12) & (0.12) & (0.33) \\ 
   & glmpath & 7 & 5 & -  & -  & 0.04 & 0.36 & -0.37 & 0.39 & -0.42 & 1.6 \\ 
   &  & (1.48) & (0) & - & - & (0.14) & (0.13) & (0.09) & (0.1) & (0.1) & (0.39) \\ 
   & hybrid glmmlasso & 6 & 5 & 0.97 & 0.82 & 0.07 & 0.99 & -1 & 1.03 & -1.04 & 0.32 \\ 
   &  & (1.48) & (0) & (0.49) & (0.44) & (0.22) & (0.28) & (0.17) & (0.22) & (0.18) & (0.21) \\ 
   & hybrid glmpath & 7 & 5 & 0.94 & 0.81 & 0.07 & 1 & -0.99 & 1.03 & -1.05 & 0.37 \\ 
   &  & (1.48) & (0) & (0.44) & (0.43) & (0.23) & (0.29) & (0.19) & (0.21) & (0.18) & (0.24) \\ 
   & thres glmmlasso & 5 & 5 & 0.98 & 0.83 & 0.07 & 0.99 & -1 & 1.04 & -1.04 & 0.26 \\ 
   &  & (0) & (0) & (0.45) & (0.42) & (0.22) & (0.29) & (0.17) & (0.22) & (0.16) & (0.21) \\ 
   & thres glmpath & 5 & 5 & 0.95 & 0.8 & 0.07 & 1 & -1 & 1.04 & -1.04 & 0.22 \\ 
   &  & (0) & (0) & (0.42) & (0.41) & (0.22) & (0.28) & (0.17) & (0.22) & (0.16) & (0.18) \\ 
   & p-glmer & 5 & 5 & 0.98 & 0.81 & 0.07 & 1 & -0.99 & 1.06 & -1.03 & 0.25 \\ 
   &  & (0) & (0) & (0.46) & (0.43) & (0.23) & (0.27) & (0.19) & (0.21) & (0.16) & (0.22) \\ 
   & p-glmmPQL & 5 & 5 & 0.97 & 0.81 & 0.06 & 0.91 & -0.93 & 0.94 & -0.96 & 0.19 \\ 
   &  & (0) & (0) & (0.43) & (0.32) & (0.2) & (0.25) & (0.18) & (0.21) &
   (0.17) & (0.13) \\ \hline

\end{tabular}
\end{center}
\end{table}
To summarize the variable selection results, we see that the thresholded
and the iterative procedures have the smallest active
sets. 
Table \ref{table12} suggests that the
covariance parameter estimates of \textit{glmmlasso} are considerable
biased whereas the covariance parameter estimates of the other
procedures are very similar. Concerning fixed-effect parameter estimation,
the two-stage approaches perform better and do not show striking
differences. Since $\beta_1$ and $\beta_2$ are not subject to penalization
(indicated by $^*$), their bias is smaller compared with the penalized
coefficients.
It can also be observed that the parameter estimates of
\textit{p-glmmPQL} are biased \citep{Jiang07}. 

\section*{Appendix D: Simulation study for the Poisson mixed model}
We are going to present some simulations where the conditional response
variable follows a Poisson distribution. It is interesting since the
behaviour is different from the binary case. Let us look at two
low-dimensional and three high-dimensional designs. Beforehand, let us sum
up the most relevant findings.

\subsection*{D.1 Summary for the Poisson mixed model}
In this subsection, we are going to give an overview over the properties of
the Poisson mixed model. We focus on the similarities and differences to
the Gaussian \citep{Schell11} as well as the binary case (Section 5 and Appendix C). 

\begin{itemize}
\item [i)] Shrinkage of the covariance parameters due to the
  $\ell_1$-penalization approach is not an issue. This is in contrast to the
  logistic mixed model and similar to the Gaussian case. 
\item [ii)] If we apply the Lasso ignoring the grouping structure within
  the observations, the recovery of the true active set fails.
\item [iii)] For Poisson mixed models, \textit{thres glmmlasso} performs
  best whereas in logistic mixed models \textit{hybrid glmmlasso}
  is preferable (see Appendix B).
\item [iv)] For the Lasso, we carried out the \texttt{R} function
  \texttt{glmpath} for a comparison. However, in all our high-dimensional
  simulation settings, the function breaks down. Hence we employ the
  \texttt{R} function \texttt{glmnet} for comparisons.
\item [v)] We observe a slow convergence rate (i.e.\ many outer iterations
  are required until convergence) in various real data applications. At the
  same time, convergence problems do occur in \texttt{glmnet}, too. The
  number of total iterations is often more than 100.
\end{itemize}

\subsection*{D.2 Low-dimensional Setting}
For the Poisson mixed models simulation study, we look at random-intercept
designs. This means that only the intercept has a random
effect. Particularly, $\bm\theta \in \mathbb{R}$ and
$\bm\Sigma_{\bm\theta}= \theta^2 \bm{1}_q$, i.e.\ $q=N$.
We present two examples in the low-dimensional setting.
\begin{itemize}
\item [$L_1$:] $N=20$, $n_C=10$, $n=200$, $p=10$, $\theta^2=1$ and $s_0=5$ with
  $\bm\beta_0=(\frac{1}{20},\frac{1}{2},-\frac{1}{2},\frac{1}{2},-\frac{1}{2},0,\ldots,0)^T$. \item
  [$L_2$:] $N=20$, $n_C=10$, $n=200$, $p=50$, $\theta^2=1$ and $s_0=5$ with $\bm\beta_0=(\frac{1}{20},\frac{1}{2},-\frac{1}{2},\frac{1}{2},-\frac{1}{2},0,\ldots,0)^T$.
\end{itemize}
The results over 100 simulation runs are shown in Table \ref{table13}.\\
\begin{table}[!h]
\footnotesize
\begin{center}
\caption{\textit{Simulation results for the Poisson mixed models $L_1$ and
    $L_2$ (median values and MADs in parentheses). A $^*$ indicates
    that the corresponding coefficient is not subject to $\ell_1$-penalization in
    the GLMMLasso$^{LA}$ estimate.}} \label{table13}
\vspace{0.2cm}
\begin{tabular}{clccccccccc}
\hline \hline
Model & Method& $|S(\hat{\bm\beta})|$ & TP & $\hat{\theta}^2$
& $\hat{\beta}_{1}^*$ & $\hat{\beta}_{2}$ &  $\hat{\beta}_{3}$ &
$\hat{\beta}_{4}$ & $\hat{\beta}_{5}$ & SE \\ \hline \vspace{3mm}
True&& 5 & 5 & 1 & 0.05 & 0.5 & -0.5 & 0.5 & -0.5 & \\
$L_1$ & oracle & 5 & 5 & 0.89 & 0.11 & 0.5 & -0.5 & 0.49 & -0.49 & 0.05 \\ 
   &  & (0) & (0) & (0.3) & (0.27) & (0.05) & (0.07) & (0.06) & (0.05) & (0.041) \\ 
   & glmmlasso & 7 & 5 & 0.88 & 0.24 & 0.44 & -0.43 & 0.42 & -0.43 & 0.1 \\ 
   &  & (1.48) & (0) & (0.31) & (0.25) & (0.05) & (0.07) & (0.06) & (0.07) & (0.089) \\ 
   & glmnet & 6 & 5 & - & 0.77 & 0.29 & -0.23 & 0.23 & -0.27 & 0.83 \\ 
   &  & (1.48) & (0) & - & (0.36) & (0.16) & (0.16) & (0.14) & (0.14) & (0.61) \\ 
   & hybrid glmmlasso & 7 & 5 & 0.9 & 0.11 & 0.5 & -0.51 & 0.5 & -0.49 & 0.064 \\ 
   &  & (1.48) & (0) & (0.3) & (0.26) & (0.05) & (0.07) & (0.07) & (0.05) & (0.053) \\ 
   & hybrid glmnet & 6 & 5 & 0.89 & 0.17 & 0.49 & -0.49 & 0.49 & -0.48 & 0.075 \\ 
   &  & (1.48) & (0) & (0.31) & (0.29) & (0.06) & (0.08) & (0.08) & (0.07) & (0.075) \\ 
   & thres glmmlasso & 5 & 5 & 0.89 & 0.11 & 0.5 & -0.5 & 0.49 & -0.5 & 0.053 \\ 
   &  & (0) & (0) & (0.3) & (0.26) & (0.05) & (0.07) & (0.06) & (0.05) & (0.043) \\ 
   & thres glmnet & 5 & 5 & 0.89 & 0.17 & 0.5 & -0.49 & 0.49 & -0.48 & 0.065 \\ 
   &  & (0) & (0) & (0.31) & (0.29) & (0.06) & (0.08) & (0.08) & (0.06) & (0.068) \\ 
   & p-glmer & 5 & 5 & 0.89 & 0.11 & 0.5 & -0.5 & 0.49 & -0.5 & 0.055 \\ 
   &  & (0) & (0) & (0.3) & (0.26) & (0.05) & (0.06) & (0.07) & (0.05) & (0.045) \\ 
   & p-glmmPQL & 5 & 5 & 0.86 & 0.14 & 0.5 & -0.5 & 0.49 & -0.49 & 0.05 \\ \vspace{3mm} 
   &  & (0) & (0) & (0.28) & (0.25) & (0.05) & (0.06) & (0.06) & (0.05) &
   (0.047) \\

$L_2$ & oracle & 5 & 5 & 0.93 & 0.06 & 0.5 & -0.49 & 0.49 & -0.49 & 0.04 \\ 
   &  & (0) & (0) & (0.38) & (0.25) & (0.06) & (0.05) & (0.04) & (0.07) & (0.041) \\ 
   & glmmlasso & 9 & 5 & 0.89 & 0.26 & 0.37 & -0.33 & 0.32 & -0.35 & 0.19 \\ 
   &  & (2.97) & (0) & (0.31) & (0.21) & (0.05) & (0.07) & (0.06) & (0.08) & (0.088) \\ 
   & glmnet & 6 & 5 & - & 0.71 & 0.21 & -0.14 & 0.12 & -0.2 & 0.96 \\ 
   &  & (2.97) & (0) & - & (0.28) & (0.15) & (0.17) & (0.17) & (0.13) & (0.58) \\ 
   & hybrid glmmlasso & 9 & 5 & 0.89 & 0.06 & 0.49 & -0.47 & 0.47 & -0.48 & 0.089 \\ 
   &  & (2.97) & (0) & (0.32) & (0.25) & (0.06) & (0.05) & (0.05) & (0.06) & (0.057) \\ 
   & hybrid glmnet & 6 & 5 & 0.85 & 0.12 & 0.48 & -0.44 & 0.43 & -0.43 & 0.24 \\ 
   &  & (2.97) & (0) & (0.33) & (0.26) & (0.1) & (0.13) & (0.13) & (0.13) & (0.28) \\ 
   & thres glmmlasso & 5 & 5 & 0.93 & 0.06 & 0.5 & -0.49 & 0.5 & -0.49 & 0.063 \\ 
   &  & (0) & (0) & (0.37) & (0.25) & (0.05) & (0.05) & (0.05) & (0.07) & (0.054) \\ 
   & thres glmnet & 5 & 5 & 0.89 & 0.13 & 0.48 & -0.45 & 0.45 & -0.44 & 0.21 \\ 
   &  & (1.48) & (0) & (0.35) & (0.27) & (0.09) & (0.14) & (0.13) & (0.14) & (0.26) \\ 
   & p-glmer & 5 & 5 & 0.93 & 0.06 & 0.51 & -0.49 & 0.49 & -0.49 & 0.048 \\ 
   &  & (0) & (0) & (0.37) & (0.24) & (0.06) & (0.05) & (0.04) & (0.07) & (0.044) \\ 
   & p-glmmPQL & 5 & 5 & 0.88 & 0.1 & 0.51 & -0.49 & 0.49 & -0.49 & 0.046 \\ 
   &  & (0) & (0) & (0.35) & (0.24) & (0.06) & (0.05) & (0.05) & (0.06) & (0.043) \\ 
   \hline
\end{tabular}
\end{center}
\end{table}
\indent We read off from the table that the thresholded methods pick less
variables than the hybrid procedures. 
The \textit{glmmlasso} covariance parameter estimates do not show a
dramatic bias here. Estimation accuracy of the fixed-effect parameters is
worst for the \textit{glmnet} based estimators.


\subsection*{D.3 High-dimensional Setting}
We examine the following high-dimensional examples:
\begin{itemize}
\item [$H_1$:] $N=40$, $n_C=10$, $n=400$, $p=500$, $\theta^2=1$ and $s_0=5$ with
  $\bm\beta_0=(\frac{1}{20},\frac{1}{2},-\frac{1}{2},\frac{1}{2},-\frac{1}{2},0,\ldots,0)^T$. \item [$H_2$:] $N=40$, $n_C=10$, $n=400$, $p=1000$, $\theta^2=1$ and $s_0=5$ with $\bm\beta_0=(\frac{1}{20},\frac{1}{2},-\frac{1}{2},\frac{1}{2},-\frac{1}{2},0,\ldots,0)^T$.
\item [$H_3$:] $N=30$, $n_C=10$, $n=300$, $p=500$, $\theta^2=0.25$ and $s_0=5$ with $\bm\beta_0=(2,\frac{1}{2},-\frac{1}{2},\frac{1}{2},-\frac{1}{2},0,\ldots,0)^T$.
\end{itemize}

The results in the form of median and rescaled median absolute deviation
(in parentheses) over 100 simulation runs are shown in Table \ref{table14}.
\begin{table}[!h]
\footnotesize
\begin{center}
\caption{\textit{Simulation results for the Poisson mixed models $H_1$,
    $H_2$ and $H_3$ (median values and MADs in
    parentheses). A $^*$ indicates that the corresponding coefficient is not
    subject to $\ell_1$-penalization in the GLMMLasso$^{LA}$
    estimate.}} \label{table14}
\vspace{0.2cm}
\begin{tabular}{clccccccccc}
\hline \hline
Model & Method& $|S(\hat{\bm\beta})|$ & TP & $\hat{\theta}^2$
& $\hat{\beta}_{1}^*$ & $\hat{\beta}_{2}$ &  $\hat{\beta}_{3}$ &
$\hat{\beta}_{4}$ & $\hat{\beta}_{5}$ & SE \\ \hline \vspace{3mm}
True&& 5 & 5 & 1 & 0.05 & 0.5 & -0.5 & 0.5 & -0.5 \\

$H_1$ & oracle & 5 & 5 & 0.94 & 0.03 & 0.5 & -0.5 & 0.5 & -0.5 & 0.02 \\ 
   &  & (0) & (0) & (0.28) & (0.17) & (0.03) & (0.03) & (0.05) & (0.04) & (0.02) \\ 
   & glmmlasso & 11 & 5 & 0.92 & 0.26 & 0.33 & -0.31 & 0.31 & -0.33 & 0.2 \\ 
   &  & (3.71) & (0) & (0.26) & (0.17) & (0.04) & (0.04) & (0.06) & (0.05) & (0.086) \\ 
   & glmnet & 6 & 5 & - & 0.8 & 0.16 & -0.12 & 0.13 & -0.18 & 1.1 \\ 
   &  & (2.97) & (0) & - & (0.26) & (0.13) & (0.15) & (0.13) & (0.12) & (0.56) \\ 
   & hybrid glmmlasso & 11 & 5 & 0.9 & 0.05 & 0.47 & -0.46 & 0.47 & -0.47 & 0.05 \\ 
   &  & (3.71) & (0) & (0.27) & (0.16) & (0.04) & (0.05) & (0.04) & (0.04) & (0.038) \\ 
   & hybrid glmnet & 6 & 5 & 0.89 & 0.13 & 0.47 & -0.45 & 0.46 & -0.47 & 0.082 \\ 
   &  & (2.97) & (0) & (0.28) & (0.23) & (0.06) & (0.09) & (0.09) & (0.07) & (0.096) \\ 
   & thres glmmlasso & 6 & 5 & 0.93 & 0.03 & 0.49 & -0.49 & 0.49 & -0.49 & 0.04 \\ 
   &  & (1.48) & (0) & (0.28) & (0.17) & (0.03) & (0.04) & (0.05) & (0.04) & (0.037) \\ 
   & thres glmnet & 5 & 5 & 0.92 & 0.12 & 0.48 & -0.47 & 0.47 & -0.49 &
   0.054 \\\vspace{3mm}    
   &  & (0) & (0) & (0.28) & (0.22) & (0.05) & (0.09) & (0.09) & (0.06) & (0.07) \\ 

$H_2$ & oracle & 5 & 5 & 0.92 & 0 & 0.5 & -0.5 & 0.5 & -0.5 & 0.017 \\ 
   &  & (0) & (0) & (0.22) & (0.16) & (0.04) & (0.03) & (0.04) & (0.04) & (0.017) \\ 
   & glmmlasso & 11 & 5 & 0.9 & 0.26 & 0.32 & -0.28 & 0.27 & -0.32 & 0.23 \\ 
   &  & (4.45) & (0) & (0.2) & (0.13) & (0.06) & (0.06) & (0.06) & (0.06) & (0.09) \\ 
   & glmnet & 7 & 5 & - & 0.76 & 0.16 & -0.09 & 0.08 & -0.14 & 1.2 \\ 
   &  & (7.41) & (0) & - & (0.2) & (0.13) & (0.13) & (0.12) & (0.15) & (0.44) \\ 
   & hybrid glmmlasso & 11 & 5 & 0.87 & 0.02 & 0.47 & -0.47 & 0.47 & -0.47 & 0.054 \\ 
   &  & (4.45) & (0) & (0.21) & (0.16) & (0.06) & (0.04) & (0.05) & (0.05) & (0.038) \\ 
   & hybrid glmnet & 7 & 5 & 0.85 & 0.09 & 0.45 & -0.45 & 0.42 & -0.45 & 0.093 \\ 
   &  & (7.41) & (0) & (0.26) & (0.14) & (0.08) & (0.11) & (0.12) & (0.08) & (0.12) \\ 
   & thres glmmlasso & 6 & 5 & 0.89 & -0.01 & 0.49 & -0.49 & 0.49 & -0.48 & 0.041 \\ 
   &  & (1.48) & (0) & (0.22) & (0.16) & (0.05) & (0.04) & (0.05) & (0.04) & (0.035) \\ 
   & thres glmnet & 5 & 5 & 0.89 & 0.08 & 0.48 & -0.47 & 0.45 & -0.46 & 0.067 \\\vspace{3mm} 
   &  & (1.48) & (0) & (0.24) & (0.14) & (0.08) & (0.12) & (0.1) & (0.08) & (0.09) \\ \vspace{3mm}

True && 5 & 5 & 0.25 & 2 & 0.5 & -0.5 & 0.5 & -0.5 \\

$H_3$ & oracle & 5 & 5 & 0.25 & 1.99 & 0.5 & -0.5 & 0.5 & -0.5 & 0.009 \\ 
   &  & (0) & (0) & (0.08) & (0.12) & (0.02) & (0.02) & (0.02) & (0.02) & (0.0096) \\ 
   & glmmlasso & 11 & 5 & 0.25 & 2.11 & 0.41 & -0.39 & 0.4 & -0.4 & 0.057 \\ 
   &  & (2.97) & (0) & (0.08) & (0.11) & (0.02) & (0.03) & (0.03) & (0.03) & (0.028) \\ 
   & glmnet & 10 & 5 & - & 2.3 & 0.33 & -0.31 & 0.3 & -0.33 & 0.25 \\ 
   &  & (5.19) & (0) & - & (0.12) & (0.05) & (0.06) & (0.06) & (0.07) & (0.12) \\ 
   & hybrid glmmlasso & 11 & 5 & 0.24 & 1.99 & 0.49 & -0.48 & 0.48 & -0.48 & 0.018 \\ 
   &  & (2.97) & (0) & (0.07) & (0.13) & (0.02) & (0.02) & (0.02) & (0.02) & (0.011) \\ 
   & hybrid glmnet & 10 & 5 & 0.23 & 1.99 & 0.49 & -0.49 & 0.49 & -0.49 & 0.016 \\ 
   &  & (5.19) & (0) & (0.08) & (0.12) & (0.02) & (0.02) & (0.02) & (0.02) & (0.012) \\ 
   & thres glmmlasso & 5 & 5 & 0.25 & 1.99 & 0.5 & -0.49 & 0.5 & -0.5 & 0.011 \\ 
   &  & (0) & (0) & (0.07) & (0.12) & (0.02) & (0.02) & (0.02) & (0.02) & (0.011) \\ 
   & thres glmnet & 5 & 5 & 0.25 & 1.99 & 0.5 & -0.5 & 0.5 & -0.5 & 0.0091 \\ 
   &  & (0) & (0) & (0.07) & (0.12) & (0.02) & (0.02) & (0.02) & (0.02) & (0.0096) \\ 
\hline
\end{tabular}
\end{center}
\end{table}

\indent 
Considering parameter estimation accuracy, the Poisson mixed model shows
that the variable screening using \textit{glmnet} fails, resulting in large
values of SE. Although the median value of TP is large, the mean value
(not shown) is below 5. This behaviour is far more obvious than in the
logistic mixed model.

We conclude by noting that the Poisson mixed model clearly shows that
it is of paramount importance to perform variable screening using
\textit{glmmlasso} and that it can not be carried out by just applying a
standard Lasso procedure (and thereby ignoring the grouping structure).


\end{document}